\documentclass[aps,pra,twocolumn, nofootinbib,amsmath,amssymb,tightenlines,floats,floatfix,showpacs]{revtex4}
\usepackage{epsfig}
\usepackage{psfig}
\usepackage{mathptm}
\usepackage[mathcal]{euscript}
\usepackage{graphicx}
\begin{document}

\title{ Differential Form of the Skornyakov--Ter-Martirosyan Equations }
\author{F. M. Pen'kov}
\email[e-mail: ]{penkov@thsun1.jinr.ru}
\email[ ]{penkov@inp.kz}
\address{Joint Institute for Nuclear Research, Dubna, Russia}
\address{Institute of Nuclear Physics, Almaty, Kazakhstan}
\author{  W. Sandhas}
\email[e-mail: ]{sandhas@physik.uni-bonn.de}
\address{ Physikalisches   Institut, Universit\"at Bonn, Bonn, Germany }
\date{\today}

\begin{abstract}

The Skornyakov--Ter-Martirosyan three-boson integral equations in
momentum space are transformed into  differential equations. This
allows us to take into account quite directly  the Danilov
condition providing self-adjointness of the underlying three-body
Hamiltonian with zero-range pair interactions. For the helium
trimer the numerical solutions of the resulting differential
equations are compared with those of the Faddeev-type AGS
equations.
\end{abstract}

\maketitle

\section{Introduction}
The discovery of a weekly bound state of two helium
atoms~\cite{2HeEx} and problems concerning the stability of Bose
condensates of alkali atoms (see, e.g.,~\cite{Fed1}) stimulated many
calculations of the properties of three-particle systems
determined by pair interactions with a large scattering length
$a_0$ compared to the range  $r_0$ of the pair forces,
\begin{equation}
\label{usl}
\frac{r_0}{a_0} \ll 1.
\end{equation}

On the one hand, this condition poses problems in the numerical
solution of the Faddeev equations with realistic atom-atom
potentials, on the other hand it forms the basis of the zero-range
model of the two-body interaction (($r_0 \to 0$)). This model
determines the motion of particles beyond the pair-interaction
region, admits an analytic simplification, and can be used for the
description of real physical systems provided the condition
(\ref{usl}) is fulfilled.

The zero-range model (ZRM) for three-body systems, though having
obvious advantages and wide application, has a considerable
drawback. The Hamiltonian of these systems is not self-adjoint
(see, for example,~\cite{Fad_DAN} ) and the Schr\"odinger equation
has quadratically integrable solutions at any energy. This fact
was pointed out in~\cite{Dan1} in analyzing poor attempts to solve
the nd-scattering problem with the use of the
Skornyakov--Ter-Martirosyan (STM) integral equations~\cite{STM}
within the zero-range model of nucleon-nucleon interaction. In
particular, at large momenta $k$ of the relative motion of a particle
and a pair the asymptotic behavior of the wave function
 was shown to have at any energy $Z$ the form
\begin{equation}
\varphi(k,Z) = A \frac{\sin(\mu_0 \ln( k)) }{k^2} +
B \frac{\cos(\mu_0 \ln( k)) }{k^2} + o(\frac{1}{k^2}).
\label{asimpt}
\end{equation}
Here, the constant $\mu_0$ depends on the ratio of the particle
masses, and only the coefficients $A$ and $B$ depend on the energy.
To determine the three-body spectrum, Danilov~\cite{Dan1}
suggested to use the relation between the coefficients $A(Z)$ and
$B(Z)$ with an energy-independent parameter $\gamma$,
\begin{equation}
A(Z)=\gamma B(Z),
\label{Dan_Cond}
\end{equation}
following from the orthogonality condition of eigenfunctions.
In the same year Minlos and Faddeev~\cite{Fad_DAN} showed that the
Danilov condition is a special case among possible  extensions of the
Hamiltonian to a self-adjoint one. Even after the extension, a solution
to the STM equation has one free parameter. This opens the
possibility  of describing real three-body
systems by adjusting the free parameter to a known spectral
point~\cite{Dan1}.

Unfortunately, the Danilov condition for an unambiguous solution
of the STM equations practically cannot be used in numerical
calculations. But it was used for an
analytic investigation of the ZRM three-body spectrum~\cite{Fad1}.
In this reference it has been shown that a three-body collapse (the
Thomas effect~\cite{Thom}) is a specific feature of the ZRM.

There are several approaches using the ZRM beyond the scope of the
STM equations, among them  the adiabatic expansion in
configuration space (see, e.g.,~\cite{Fedor}). Just in this approach
the so-called Efimov effect was
observed, i.e., the fact that three-body spectra concentrate on
zero total energy if the two-body scattering length $|a_0|$ tends
to infinity~\cite{Efim1}.
The result of adiabatic expansions is an infinite system of
differential equations coupled in terms of first derivatives. The
problem of non-self-adjointness of the three-body Hamiltonian is
solved by cutting off the effective interaction at small distances.
This method of regularization implicitly introduces three-body
forces. In this case, the cut-off radius   plays the role of the
free parameter.

Three-body forces are introduced more explicitly
 in the effective field theory (EFT)~\cite{EFT},
causing an integral equation that is similar to the STM equation,
but contains  artificial terms. A free parameter enters the
phenomenological terms of the kernel of the integral equation and
the free terms of the integral equation for the scattering
problem~\cite{EFT}.

Finally, we would like to draw attention to the two-pole
$t$-matrix model for the description of three-boson
systems~\cite{Pen2003} which employs the
Alt-Grassberger-Sandhas (AGS)~\cite{AGS} version of the Faddeev
equations~\cite{Fad_eq}. The position of the second pole on the
unphysical sheet is treated in this model as a parameter of the integral
equation. When the position of the second pole tends to infinity,
the STM equations are reproduced~\cite{Pen2003}. This method deals
with compact equations and is well suited for numerical
calculations, but is of little use for analytic considerations.

\section{Formalism}
In the present paper we follow another strategy.
Transforming the STM equations into differential equations
allows us to take into account the Danilov condition quite directly.
For this purpose, we construct an infinite system of differential equations in
 momentum space with a very simple relation between the equations.


The homogeneous part of the STM equation for the elastic scattering
amplitude~\cite{STM} of a boson of mass $m$ on a two-boson bound state
of energy $\varepsilon=-\varkappa^2/m$ can be represented in a form
convenient for further analysis,
\begin{equation}
F(k_i)=\frac{2}{\pi}\int \limits_0^{\infty}
\ln \frac{k_i^2+k^2+k_i k+ \lambda^2 }{k_i^2+k^2-k_i k+ \lambda^2}
\frac{F(k)\ d k}{\sqrt{\lambda^2+k^2 3/4}-\varkappa} ,
\label{IntEq}
\end{equation}
where $\lambda^2=-m Z$.
The function $F(k)$  is related to the wave function of the
three-boson system $\varphi(k)$ via
$F(k)=k (\sqrt{-mZ+k^2 3/4}-\varkappa)\varphi(k)$.
It should be noted that Eq.~(\ref{IntEq}) implies
$\ F(-k)=-F(k)$.

For further transformations of~(\ref{IntEq}) it is convenient,
following~\cite{Fad1}, to substitute the variables
$k=\frac{\lambda (t^2-1)}{\sqrt{3}t}$ ($F(k(t))\equiv F(t)$)
 and use the Mellin transformation
 ${\cal F}(s)= \int_0^{\infty}t^{s-1}F(t) dt $
\begin{equation}
{\cal F}(s)=L(s){\cal F}(s) +
L(s) 2\frac{\varkappa}{\lambda} \int \limits_0^{\infty}dt
\frac{t^{s-1}}{\frac{t^2+1}{t}-2\frac{\varkappa}{\lambda}} F(t),
\label{EqFs}
\end{equation}
where
$$
L(s)=\frac{8}{\sqrt{3}}
\frac{\sin(\frac{\pi}{6}s)}{s \ \cos(\frac{\pi}{2}s)}.
$$
By inverting  $(1-L(s))$,  Eq.(\ref{EqFs}) goes over into
\begin{equation}
{\cal F}(s)= {\cal F}_{as}(s) +
 \frac{ L(s)}{1-L(s)} 2\frac{\varkappa}{\lambda} \int
\limits_0^{\infty}dt
\frac{t^{s-1}}{\frac{t^2+1}{t}-2\frac{\varkappa}{\lambda}} F(t).
\label{F_inv}
\end{equation}
 Here, the general solution of
\begin{equation}
(1-L(s)){\cal F}_{as}(s)=0
\label{Fas}
\end{equation}
for $s=\pm i\mu_0$ ($\mu_0=1.00623...$) had to be added.
We recall that Eq.(\ref{Fas})
 was used by Minlos and Faddeev~\cite{Fad1} to  study the STM
equation spectra. In the variable $t$ the solution is
$$
F_{as}(t) \propto \sin(\mu_0 \ln t ).
$$

When carrying out the backward Mellin transformation we see that the function $F$ can
be written as a sum
\begin{equation}
F=F_0+\sum_{i=1}^{\infty} F_i
\label{SUM}
\end{equation}
over the
residues of the function $(1-L(s))^{-1}$. The term
$F_0$ is generated by the poles at the points $s=\pm i\mu_0$ and
the terms $F_i$ correspond to the real positive solutions of the
equation $1-L(X_i)=0$. It can be shown that the
components of the function $F$ satisfy the differential equations
\begin{eqnarray}
\label{F0} t(\frac{d}{d t} t \frac{d}{dt}F_0)+ \mu_0^2 F_0 +
\frac{2 \mu_0}{D(\mu_0)}
\frac{2\frac{\varkappa}{\lambda}}{\frac{t^2+1}{t}-2\frac{\varkappa}
{\lambda}}F(t)=0, \\
\label{Fi} t(\frac{d}{d t} t \frac{d}{dt}F_i) - X_i^2 F_i +
\frac{2 X_i}{{D(X_i)}}
\frac{2\frac{\varkappa}{\lambda}}{\frac{t^2+1}{t}-2\frac{\varkappa}
{\lambda}}F(t)=0,
\end{eqnarray}
where $ D(X_i)=(1-L(s))'|_{s=X_i} $ and $
iD(\mu_0)=(1-L(s))'|_{s=i\mu_0} $. Boundary conditions at $t=1$,
($k=0$) follow from the above-mentioned  antisymmetry of $F(k)$, i.e., $F_i(t)|_{t=1}=0; \ i=0,1,...$. At
large values of the argument one can use either the property of
boundedness, or the asymptotic form following from
condition~(\ref{asimpt}): $F(t)|_{t \to \infty} \to \sin(\mu_0\ln
t + \delta)$. Comparing the Danilov condition~(\ref{Dan_Cond})
with this asymptotic form of $F$ we get an extra condition for the
spectrum
\begin{equation}
\delta + \mu_0\ln(\sqrt{3}/\lambda)= const(Z) +\pi n.
\label{spetr_cond}
\end{equation}

It can be verified that the asymptotics of $F$ coincides with that of
$F_0$ and the contribution of the remaining terms decreases at
infinity. At $\varkappa/\lambda=0$  the system of
equations~(\ref{F0},\ref{Fi}) becomes uncoupled and has a simple
solution, $F=F_0=\sin(\mu_0 \ln(t))$, which gives, together with
condition~(\ref{spetr_cond}), the spectrum $E_n =E_0 \exp(-2\pi
n)$. This spectrum  contains the tree-body collapse
at large $\lambda$ for $n<0$~\cite{Fad1},
and  Efimov's concentration towards the
point $Z=0$  at extremely small $\varkappa$ for $n>0$. At small
finite $\varkappa$ we can restrict ourselves
to Eq.~(\ref{F0}) that belongs to a
 well known class of differential equations (Heun)~\cite{Heun}.
%
\begin{figure}[t]
\parbox{8.1cm}{
\mbox{\epsfig{file=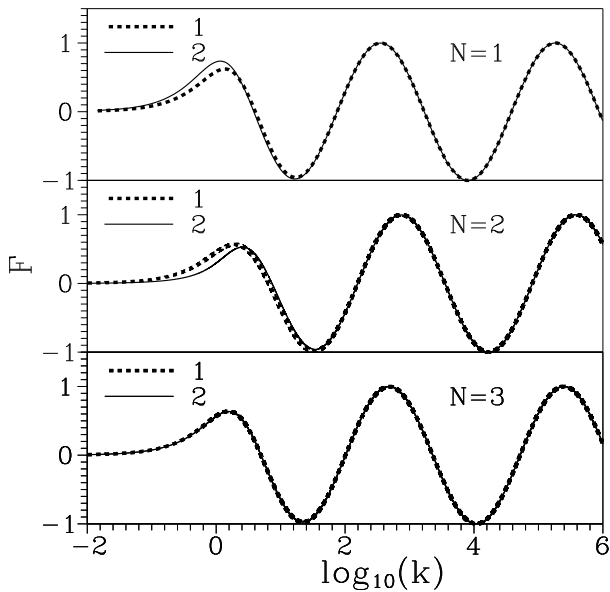,width=8cm}}
\caption { Convergence of
the solutions of Eqs.~(\ref{F0},\ref{Fi}) for increasing $N$:
Curve 1  shows $F_L$, curve 2 shows $F_R$ }
\label{Fig1}
}
\end{figure}
\begin{figure}[t]
%
\hfill
\parbox{8.1cm}{
\mbox{\epsfig{file=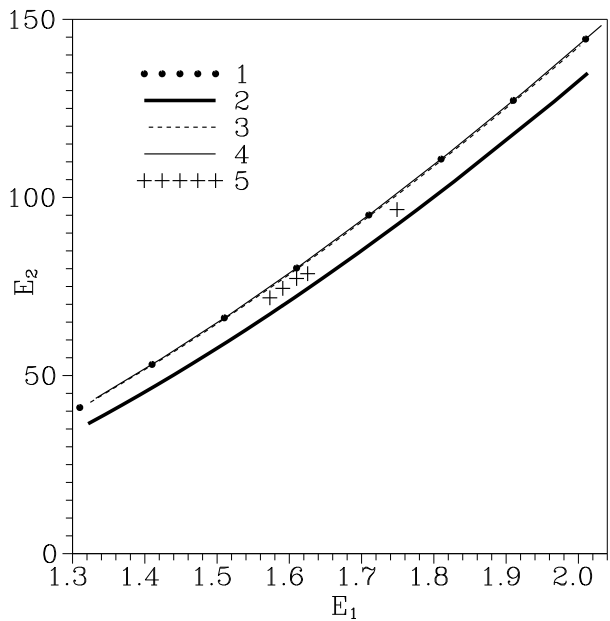,width=8cm}}
\caption { Trimer binding energies: {\bf 1} calculations by
Eqs.~(\ref{F0},\ref{Fi}); {\bf 2} calculations~\cite{Pen2003};
{\bf 3}, {\bf 4}  calculations
for  three- and four-level trimers;
{\bf 5}  calculations~\cite{RouSaf,Mot} for a helium
trimer~\cite{RouSaf,Mot} } \label{Fig2} }
\end{figure}

\section{Results}
Let us first
demonstrate the convergence of the solutions for an increasing  number $N$
of equations~(\ref{F0},\ref{Fi}).
To this end, we denote the
solution of this system of equations  by $F_R$ which, upon substitution
into the right-hand side of~(\ref{IntEq}), gives the function
$F_L$. The degree of proximity of $F_R$ and $F_L$ shows to what
extent the solution of the system~(\ref{F0},\ref{Fi}) is close to
that of the integral equation.
Figure~\ref{Fig1}  shows this convergence for $N=1,2,3$.
The energy parameter $Z/\varepsilon =(\lambda/\varkappa)^2$
was chosen to be  1.57. This  value corresponds
to the calculations~\cite{RouSaf} for the binding energy of a
helium trimer. In this way we fix the free parameter characteristically
occurring in all STM treatments.
The good convergence achieved already for $N=3$ indicates the
efficiency of our differential equations approach.

Instead of making a similar consideration for the bound-state spectrum, we compare
our present results with alternative calculations.
In this context we use the binding energies
of a helium trimer, obtained in~\cite{RouSaf,Mot} for realistic pair potentials,  and
the calculations of the binding energies via  Faddeev-type AGS integral
equations in the framework of the above-mentioned two-pole pair
$t$-matrix~\cite{Pen2003}. Since, depending on the position of the
$t$-matrix pole  on the unphysical sheet, one can obtain as many
bound trimer states as one wishes, we will label the highest bound
state by  "1" and the following one by "2".

Figure~\ref{Fig2} shows energy $E_2$  as a function of  energy  $E_1$ calculated with
Eqs.~(\ref{F0},\ref{Fi}) for $N=21$.  Also shown are the
calculations for a two-level trimer within the two-pole pair
$t$-matrix~\cite{Pen2003}, and our calculations in the same model
when the parameters of the pair $t$-matrix admit the existence of
three and four bound states of a boson trimer. All the energies
are given in units of the dimer  energy.
We see that there is only a rather small
difference between the corresponding curves.
 Quite interesting is the fact that the
 calculations for a helium trimer~\cite{RouSaf,Mot} with realistic pair
potentials lie just between these curves.

Thus, we have demonstrated that Eqs.~(\ref{F0},\ref{Fi}) represent a very efficient tool
for calculating three-boson trimers.
These equations can also be extended to scattering problems.

The authors are grateful to A.K. Motovilov and E.A. Kolganova for
fruitful discussions. This work was supported by the Deutsche
Forschungsgemeinschaft (DFG) and the Russian Foundation for Basic
Research.

\end{document}